\documentclass[pre,showpacs]{revtex4}
\usepackage{amssymb}
\usepackage[dvips]{graphicx}
\usepackage{amsmath}
\usepackage{graphicx}
\usepackage{makeidx}
\usepackage{subfigure}
\begin{document}
\title{Self-organized criticality in an interface-growth model with quenched randomness}
\author{Hidetsugu Sakaguchi}
\affiliation{Department of Applied Science for Electronics and Materials,
Interdisciplinary Graduate School of Engineering Sciences, Kyushu
University, Kasuga, Fukuoka 816-8580, Japan}
\begin{abstract} 
We study a modified model of the Kardar-Paris-Zhang equation with quenched disorder, in which the driving force decreases as the interface rises up. A critical state is self-organized, and the anomalous scaling law with roughness exponent $\alpha\sim 0.63$ is numerically obtained.  
\end{abstract}
\maketitle
Nonequilibrium interface growth and its scaling properties have intensively been studied.~\cite{rf:1} There are a number of applications of  random 
interface-growth problems in molecular beam epitaxy, bacteria-colony growth, and fluid invasion in porous media. 
In many rough interfaces, the root-mean-square (rms) width $W(l,t)$ obeys a dynamic scaling law:
\begin{equation}
W(l,t)\equiv\langle(h(x,t)-\langle h(x,t)\rangle)^2\rangle^{1/2}\sim l^{\alpha}\tilde{h}(t/l^z),
\end{equation}
where $h(x,t)$ is the surface height at time $t$, $z=\alpha/\beta$ is a dynamic exponent, $l$ is an interval size, and the scaling function $\tilde{h}(u)$ satisfies asymptotically $\tilde{h}(u)\sim u^{\beta}$ for $u\ll 1$ and $\tilde{h}(u)\rightarrow 1$ for $u\gg 1$. 
The exponent $\alpha$ is called the roughness exponent and $\beta$ is the growth exponent.  
It has been widely believed that many systems lie in the same universality class as the Kardar-Parisi-Zhang (KPZ) equation.~\cite{rf:2} 
The $(1+1)$ KPZ equation has a form:
\begin{equation}
\frac{\partial h}{\partial t}=\nu \frac{\partial^2h}{\partial x^2}+\lambda\left(\frac{\partial h}{\partial x}\right )^2+\eta(x,t),
\end{equation}
where $\nu$ and $\lambda$ are constants, and $\eta(x,t)$ denotes the Gaussian white noise satisfying $\langle \eta\rangle=0$ and $\langle \eta(x,t)\eta(x^{\prime},t^{\prime})\rangle=2D\delta(x-x^{\prime})\delta(t-t^{\prime})$. 
The roughness exponent $\alpha$ is $1/2$ and the growth exponent $\beta=1/3$ in the $(1+1)$ KPZ equation. However, many experiments give larger values of $\alpha$ in the range 0.6-0.9~\cite{rf:3,rf:4,rf:5,rf:6}. 
To explain the anomalous exponent, some authors studied modified models of the KPZ equation, in which the noise was assumed to have a power-law distribution,~\cite{rf:7} although the physical origin of such a noise remains unclear. 
The randomness is quenched in space in many experiments,  Thus, the KPZ equation with quenched disorder (KPZQ equation) was also studied theoretically.~\cite{rf:8,rf:9} 
The KPZQ equation in one dimension is written as 
\begin{equation}
\frac{\partial h}{\partial t}=\nu \frac{\partial^2h}{\partial x^2}+\lambda\left(\frac{\partial h}{\partial x}\right )^2+f+\eta(x,h),
\end{equation}
where $\eta(x,h)$ denotes the quenched disorder satisfying $\langle \eta\rangle=0$ and $\langle \eta(x,h)\eta(x^{\prime},h^{\prime})\rangle=2D\delta(x-x^{\prime})\delta(h-h^{\prime})$. The nonlinear coefficient $\lambda$ is proportional to the average velocity $v$ from the kinematic origin in most case, however, such a nonlinear term appears in anisotropic media even for $v=0$.~\cite{rf:10} 

The KPZQ equation exhibits a pinning transition when the driving force $f$ is decreased. When the driving force $f$ is larger than a critical value $f_c$, 
the interface grows with an average velocity. When $f$ is below $f_c$, the interface is pinned in the quenched random medium. 
At the critical driving force $f_c$, the scaling law $W(l)\sim l^{\alpha}$ is satisfied, and the exponent $\alpha$ is evaluated at $\alpha\sim 0.63$. The exponent $\alpha$ is related to two exponents $\nu_{\|}$ and $\nu_{\bot}$ in the directed percolation problem as $\alpha=\nu_{\|}/\nu_{\bot}$, where $\nu_{\|}$ is the exponent for the longitudinal correlation length $\xi_{|}$ and $\nu_{\bot}$ is the exponent for the transverse correlation length $\xi_{\bot}$.~\cite{rf:11} The exponent $\alpha=0.63$ is close to the experimental exponents in paper wetting~\cite{rf:11}. The scaling law is theoretically satisfied only at the critical driving force $f_c$. However, the scaling law is observed even if the parameter is not precisely controlled to be the critical value.  The critical condition might be to be self-organized by a certain unclear mechanism. An interface-growth model which exhibits the self-organized criticality (SOC) was proposed by Sneppen~\cite{rf:12}. In his algorithm (the model B in [12]), the site with the smallest random force $\eta(x,h)$ is selected , and the updating $h\rightarrow h+1$ is done at the site.  Then, neighboring sites are sequentially updated as $h\rightarrow h+1$, if $|h(x)-h(x-1)|\le 1$ is satisfied. 
Such a chain reaction rule is used in many models exhibiting the self-organized criticality.~\cite{rf:13}
A self-organized random interface with $\alpha=0.63$ was obtained in the model of Sneppen, but the physical origin of the updating rule is not clear.  Thus, we think that the problem of the anomalous exponent in random interfaces is not completely solved yet. In this brief report, we propose another simple mechanism of the self-organized criticality in a modified model of the KPZQ equation. 

We consider an experiment of paper wetting as performed in [11]. The driving force in this problem is the surface tension $F$. The paper is dipped into a basin filled with suspensions of ink or coffee. When the interface rises up, the upward driving force decreases owing to the gravity as $f=F-d_0h$, where $d_0$ is proportional to the density $\rho$ multiplied by the gravitational acceleration $g$,  because the gravitational potential increases with the square of the interface height.~\cite{rf:14}. Furthermore, the evaporation from the surface of the paper needs to be taken into consideration, if the humidity is low. As a rough approximation, the interface height decreases on the average by $-e h$, because the volume of the liquid is proportional to $h L w$ where $w$ is the average thickness and $L$ is the side length of the paper, the evaporation rate is proportional to the area $2h L$ of two surfaces of the paper, and the evaporation rate is proportional to the decrease rate of the liquid volume. That is, the evaporation effect also decreases the upward driving force as $f=F-eh$ as a first order approximation.  
Thus, our model equation is written as 
\begin{equation}
\frac{\partial h}{\partial t}=\nu \frac{\partial^2h}{\partial x^2}+\lambda\left(\frac{\partial h}{\partial x}\right )^2+F-d h+\eta(x,h),
\end{equation}
where $d=d_0+e$. 
If the term $dh$ is approximated at the spatial average $d\langle h\rangle$, 
 where $\langle h(t)\rangle$ is the spatial average of the interface height in system size $L$: $\langle h(t)\rangle=(1/L)\int_0^Lh(x,t)dx$,
our model equation is expressed as 
\begin{equation}
\frac{\partial h}{\partial t}=\nu \frac{\partial^2h}{\partial x^2}+\lambda\left(\frac{\partial h}{\partial x}\right )^2+F-d\langle h\rangle+\eta(x,h).
\end{equation}
\begin{figure}[tbp]
\begin{center}
\includegraphics[height=3.5cm]{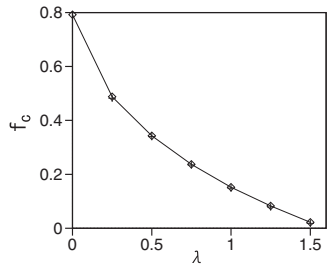}
\end{center}
\caption{Critical driving force $f_c$ (solid line with rhombi) in the discrete KPZQ equation (6) at $\nu=5$ and $L=10000$ for $\lambda=0,0.25,0.5,0.75,1,1.25$ and 1.5. The marks $+$ denote the average values of $F-d\langle h\rangle$ for the same values of $\lambda$ in the time evolution of Eq.~(7).}
\label{f1}
\end{figure}
This model equation is closely related to the KPZQ equation, or it can be interpreted as a modified KPZQ equation with a feedback term. 
If $F>f_c$, the interface grows upward from the flat initial condition $h(x,0)=0$. The driving force becomes weaker, as the average height $\langle h\rangle$ increases. 
The interface growth is expected to stop when $F-d\langle h\rangle\simeq f_c$ is satisfied owing to the pinning transition in the KPZQ equation. 
After the interface is pinned, the critical condition $F-d\langle h\rangle\simeq f_c$ is maintained forever, because  $\langle h\rangle$ is constant in time. Thus, the self-organized criticality can be realized by this very simple mechanism. 

We have performed direct numerical simulations of Eq.~(3) and (5) using a simple Euler method. The KPZQ equation (3) is discretized as  
\begin{equation}
h(x,t+\Delta t)=h(x,t)+\Delta t \left [\nu\{h(x+1,t)-2h(x,t)+h(x-1,t)\}+\lambda\{h(x+1,t)-h(x-1,t)\}^2/4+f+\eta(x,h)\right ].
\end{equation}
The modified model Eq.~(5) is discretized as 
\begin{equation}
h(x,t+\Delta t)=h(x,t)+\Delta t \left [\nu\{h(x+1,t)-2h(x,t)+h(x-1,t)\}+\lambda\{h(x+1,t)-h(x-1,t)\}^2/4+F-d\langle h\rangle+\eta(x,h)\right ].
\end{equation}
The parameters are fixed to be $\nu=5$ and $\Delta t=0.01$. We have used quenched disorder $\eta(x,h)$ which is uniformly distributed between 0 and 3.~\cite{rf:9}
\begin{figure}[tbp]
\begin{center}
\includegraphics[height=3.5cm]{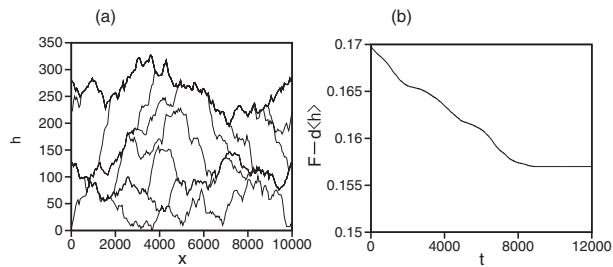}
\end{center}
\caption{(a) Successive snapshot profiles of $h(x,t)$ at $1000\times n$ ($n=1,2,\cdots,12$) by Eq.~(7) at $\nu=5,\lambda=1,F=0.17,d=0.00005$ and $L=10000$.  (b) Time evolution of $F-d\langle h\rangle$ for the numerical simulation shown in (a).}
\label{f2}
\end{figure}
\begin{figure}[tbp]
\begin{center}
\includegraphics[height=3.5cm]{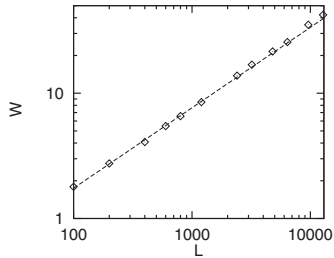}
\end{center}
\caption{Relation of $W(L)$ as a function of $L$ after the interfaces are pinned for Eq.~(7) at $\nu=5$, $\lambda=1, F=0.17$ and $d=0.00005$. 
The dashed line is the fitting curve $W(L)=0.0907L^{0.642}$ by the least squares method.}
\label{f3}
\end{figure}
\begin{figure}[tbp]
\begin{center}
\includegraphics[height=3.5cm]{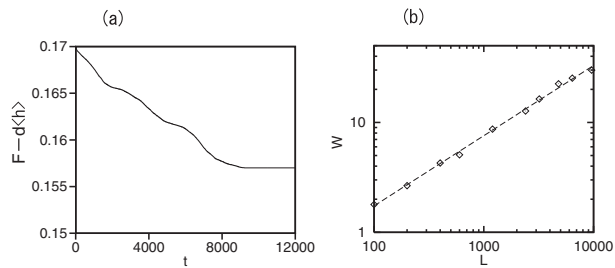}
\end{center}
\caption{(a) Time evolution of $F-d\langle h\rangle$ by the discretized model of Eq.~(4) at $\nu=5,\lambda=1,F=0.17,d=0.00005$ and $L=10000$.  (b) Relation of $W(L)$ as a function of $L$ after the interfaces are pinned for Eq.~(4) at $\nu=5$, $\lambda=1, F=0.17$ and $d=0.00005$. 
The dashed line is the fitting curve $W(L)=0.0917L^{0.639}$ by the least squares method}
\label{f2}
\end{figure}
Figure 1 shows a critical curve $f_c$ (solid line) of the discrete KPZQ equation (6) as a function of $\lambda$. The system size is $L=10000$ and periodic boundary conditions $h(L,t)=h(0,t)$ are assumed in the numerical simulation. The pinning phenomenon is observed below the critical curve. 
The critical driving force $f_c$ decreases with $\lambda$. 

Figure 2(a) shows the time evolution of $h(x,t)$ for $L=10000$, $\lambda=1,d=0.00005$ and $F=0.17$ by Eq.~(7). Snapshot profiles of $h(x,t)$ at $1000\times n$ ($n=1,2,\cdots, 12$) are shown in Fig.~2(a). 
The initial condition is $h(x,0)=0$. For small $t$, the interface rises up randomly. As the interface rises up, the driving force becomes weak and finally the interface is pinned. Near the pinning transition, the interface growth occurs intermittently both in space and time, that is, the interface is pinned in most positions and the interface growth occurs locally.  
 Figure.~2(b) shows the time evolution of $F-d\langle h\rangle$ for the numerical simulation shown in Fig.~2(a).  The driving force $F-d\langle h\rangle$ decreases and finally takes a constant value by the pinning. The final value is 0.157 in this quenched disorder.  
The final values slightly depend on the quenched disorder. 
We have calculated the average value using 50 random quenched media.  
The average values of $F-d\langle h\rangle$ at the stationary state for various values of $\lambda$ are plotted by $+$ marks in Fig.~1. The stationary values of $F-d\langle h\rangle$ almost locate on the critical curve $f_c(\lambda)$ of the KPZQ equation. This implies that the self-organized criticality is realized in our model.  We have calculated numerically the rms width $W(L)$ after the interfaces are pinned for various values of $L$ at $\lambda=1, F=0.17$ and $d=0.00005$. Figure 3 shows a relation of $W(L)$ as a function of the system size $L$.  The exponent $\alpha$ is evaluated by the least squares method as $\alpha=0.64$, which is close to 0.63 for the KPZQ equation at the critical point.  The anomalous critical exponent is realized at the stationary state in our model, even if the initial driving force $F$ is not precisely controlled to be $f_c$.   

To check the validity of the mean-field approximation (5) to Eq.~(4), we have performed direct numerical simulation of the discretized model of Eq.~(4) with $d=0.00005$. 
Figure 4(a) shows the time evolution of $F-d\langle h\rangle$ at $\nu=5,\lambda=1,F=0.17$ and $L=10000$. The time evolution is very close to the one shown in Fig.~2(b). Figure 4(b) shows the relation of $W(L)$ vs. $L$.  The exponent $\alpha$ is evaluated by the least squares method as $\alpha=0.64$.  These numerical results show that the approximation by Eq.~(5) is rather good in our simulation at sufficiently small $d$. However, it is expected that the scaling properties on very large length scales in Eq.~(4) are different from the ones with the roughness exponent $\alpha=0.63$.  The crossover length is expected to be O($\sqrt{\nu/d}$), above which the linear damping term $-dh$ in Eq.~(4)  becomes dominant compared to the diffusion term $\nu\partial^2h/\partial x^2$.  On the other hand, in the mean-field model (5), the damping term acts only for the  spatial average $\langle h\rangle$, and does not have a direct effect for the spatial fluctuation of $h(x,t)$.

To summarize, we have found a very simple mechanism of self-organized criticality in a modified model of the KPZQ equation. The critical pinning state is naturally realized by decreasing the driving force.  Both the gravitational effect and the evaporation effect are considered to decrease of the driving force effectively. The gravitational effect is controlled by changing the inclination angle of the paper and the evaporation effect is controlled by the humidity.   
This might be a mechanism of the anomalous scaling in paper wetting experiments. Similar mechanisms might work even for other experiments of interface growth in quenched random media. For example, there is another feedback mechanism for the self-organized criticality in the paper wetting problem where the paper is set horizontally and the gravitational effect does not work, and the evaporation effect is neglected. The average growth velocity $v$ of the interface decreases as $t^{-1/2}$ (Washburn's law)  because of the viscous resistance force in proportion to $\langle h\rangle$.~\cite{rf:14,rf:15} Then, the $\lambda$ parameter decreases effectively in time owing to the kinematic effect, and the self-organized critical pinned state will be realized,  because the critical driving force $f_c$ is a decreasing function of $\lambda$ in the KPZQ equation as shown in Fig.~1.  Furthermore, in general, there might exist other systems, in which critical states are self-organized by some feedback terms.  We would like to study such self-organized criticality by the feedback effect in the future.

\end{document}